\documentclass{elsart}

\usepackage{natbib}

\usepackage{epsfig}

\usepackage{amssymb}

\begin{document}

\newcommand{\chandra}{{\it Chandra}}
\newcommand{\snr}{3C~58}
\newcommand{\psr}{PSR~B1509--58}

\def\la{\ifmmode\stackrel{<}{_{\sim}}\else$\stackrel{<}{_{\sim}}$\fi}
\def\ga{\ifmmode\stackrel{>}{_{\sim}}\else$\stackrel{>}{_{\sim}}$\fi}
\def\farcm{\hbox{$.\mkern-4mu^\prime$}}
\def\farcs{\hbox{$.\!\!^{\prime\prime}$}}

\begin{frontmatter}



\title{
The Devil is in the Details: Compact Structures in Pulsar Wind Nebulae
}


\author{Patrick Slane}

\address{Harvard-Smithsonian Center for Astrophysics}
\address{60 Garden Street, Cambridge, MA 02138, USA}

\begin{abstract}

The large-scale structure of pulsar wind nebulae (PWNe) tells us a 
considerable amount about their average magnetic fields, the total particle 
input from the pulsar winds, and the confining pressure at their outer 
boundaries. However, the details of the pulsar outflow, the sites of 
shocks and particle acceleration, the effects of instabilities in the 
magnetic field, and the interaction between the relativistic wind and 
the surrounding ejecta are contained in small-scale structures, where we 
observe jets and toroidal structures, time-varying emission from compact 
clumps, and filaments in both the inner and outer regions of the nebulae. 
Here I review recent observational studies of compact structures in PWNe 
and present current scenarios (and questions) regarding their origin.

\end{abstract}


\end{frontmatter}

\section{Introduction}

Our basic understanding of PWNe stems from the picture presented
by \cite{rg74}, and expanded upon by \cite{kc84},
in which an energetic axisymmetric wind is injected from a pulsar into 
its surroundings. As illustrated schematically in Figure 1, the 
structure of a PWN is regulated by the input power from the pulsar and the 
density of the medium into which the nebula expands; the pulsar wind and
wound-up toroidal magnetic field inflates a bubble which is confined in the 
outer regions by the expanding shell of ejecta or interstellar material 
swept up by the SNR blast wave. The boundary condition established by the 
expansion at the nebula radius $r_N$ results in the formation of a wind 
termination shock at which the highly relativistic pulsar wind is decelerated 
to merge with the particle flow in the nebula. The shock forms at the radius 
$r_w$ at which the ram pressure of the wind is balanced by the internal 
pressure of the PWN:
\begin{equation}
r_w = \sqrt{\dot E/(4 \pi \eta c p)},
\end{equation}
where $\dot E$ is the rate at which the pulsar injects energy into the wind, 
$\eta$ is the fraction of a spherical surface covered by the wind, and $p$ is 
the total pressure outside the shock.  Ultimately, the pressure in the nebula 
is believed to reach the equipartition value; a reasonable pressure estimate 
can be obtained by integrating the broad-band spectrum of the nebula, using 
standard synchrotron emission expressions, and assuming equipartition between 
particles and the magnetic field. Typical values yield termination shock 
radii of order 0.1~pc, which yields an angular size of several arcsec at 
distances of a few kpc. 

\begin{figure}
\centerline{\epsfig{file=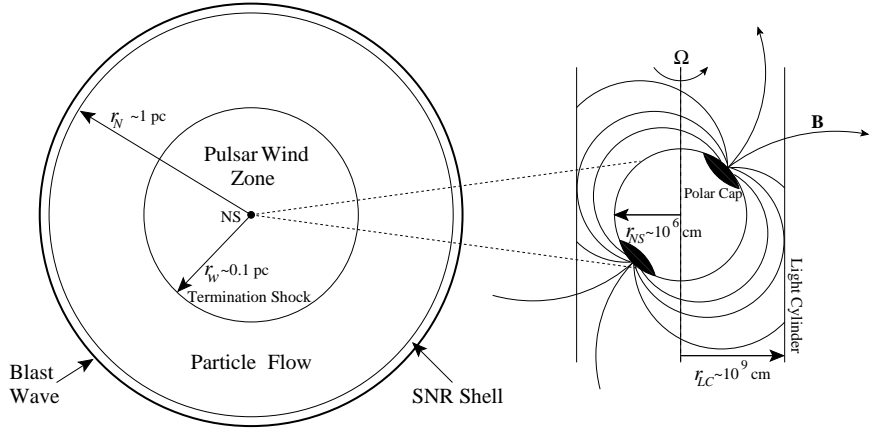,width=4.5in}}
\caption{
Schematic view of a pulsar and its wind nebula. See the text for
a complete description. (Note the logarithmic
size scaling in the PWN figure when comparing with images shown elsewhere in
the text.)
}
\end{figure}

As the relativistic fluid comprising the PWN encounters the freely-expanding
ejecta, Rayleigh-Taylor instabilities result in the formation of a network
of dense, optical line-emitting filaments 
\citep{jun98}.
The
density and magnetic field strength become enhanced in regions where the
PWN creates these filaments, producing enhanced synchrotron emission observed
as radio filaments. 
Due to the pinching effect of the global toroidal magnetic field, the overall 
morphology of a young PWN is often elongated along the pulsar spin axis
(Begelman \& Li, 1992\nocite{bli92}; van der Swaluw et al., 
2004\nocite{vdk04}).
Along the rotation axis the flow becomes collimated, producing jets. 
Pinch instabilities may disrupt the toroidal structure, however, changing the 
structure of the magnetic field in the outer nebula regions and relaxing the
collimation of the jets far from the pulsar \citep{beg98}.

The overall geometry of the PWN, as well as that of the emission from jets
or ring-like structures near the termination shock, thus provides a direct
indication of the pulsar geometry. The details of the jet morphology and the
emission structure in the postshock region provide the strongest constraints
available on wind composition and particle acceleration in PWNe.
For cases in which the pulsar proper
motion is also known, constraints on the kick velocity mechanism can be
derived based on the degree of alignment between the velocity vector and
the pulsar spin axis.

\section{Pulsar Jets and Tori}

In the inner portions of the Crab Nebula, optical wisps mark the position of
the wind termination shock, at a distance of $\sim 0.1$~pc from the pulsar. 
The brightness and position of these wisps varies
in time, with inferred outflow speeds up to $0.7 c$ 
\citep{hes98}.
As shown in Figure 2 (left),
high resolution X-ray images reveal a ring of emission at 
the position of the wisps \citep{wht+00}, 
providing a direct 
connection between the unshocked pulsar wind and the bulk properties of 
the nebula. Material from the inner ring forms a series of toroidal X-ray 
wisps that are variable with time \citep{hmb+02}. 
The geometry of these
X-ray features imply a tilted torus, and a jet of material flows perpendicular
to the plane of the toroid, extending some 0.25~pc from the pulsar. A faint 
counterjet is also observed, along with significantly enhanced X-ray 
emission from the leading portion of
the toroid. Both effects presumably result from Doppler beaming of the
outflowing material, whereby the X-ray intensity varies with viewing angle:
\begin{equation}
\frac{I}{I_0} = \left[\frac{\sqrt{1 - \beta^2}}{1 - \beta \cos\theta}
\right]^{\alpha +1}
\end{equation}
where $\beta = v/c$ is the flow speed, $\alpha$ is the photon index of
the synchrotron spectrum, and $\theta$ is the angle of photon propagation
relative to the flow direction \citep{pell87}.
One troubling aspect of this suggestion is that the
brightness distribution around the inner ring does not match that of the
outer toroid; indeed, the brightness is rather uniform except for some 
small clump-like structures that vary in position and brightness with time
(see Section 3). 

The process by which particles are accelerated from $\gamma \sim 10^4$, as
expected in the wind outside the pulsar light cylinder, to the $\gamma
> 10^6$ required to explain the emission of synchrotron X-rays (assuming
a magnetic field of $\sim 300 \mu{\rm G}$), is not understood.
Moreover, the production of radio-emitting electrons is not even predicted
by the model of Kennel and Coroniti (1984). Yet recent VLA observations
show variable structures very similar to the optical wisps, indicating
that acceleration of the associated particles must be occurring in the same
region as for the X-ray--emitting particles 
\citep{bhf+04}. 

\begin{figure}
\centerline{\epsfig{file=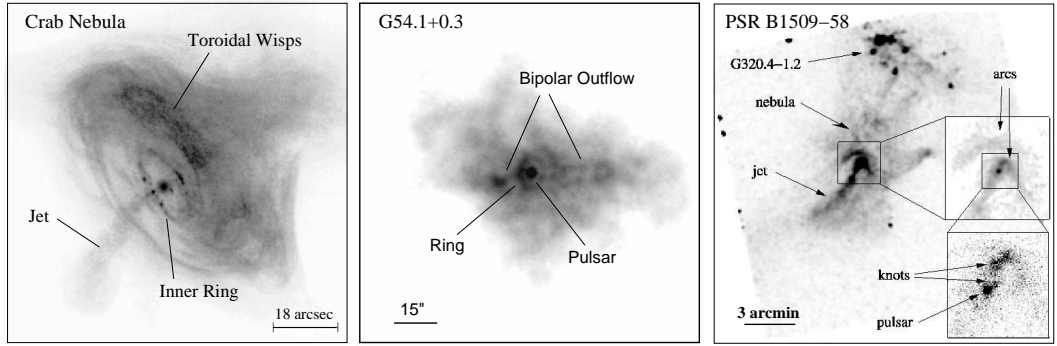,width=5.5in}}
\caption{
{\it Chandra} images of the Crab Nebula (left), G54.1+0.3 (center), and
PSR~B1509$-$58 (right) showing the complex emission 
from these PWNe, including jet outflows and toroidal structures.
}
\end{figure}

A handful of other PWNe display X-ray features that suggest the presence of 
extended ring-like structures and narrow collimated components.
The size of the ring-like features (or, in some cases, the slight
extent of the compact source) places the emission region near the
pulsar wind termination shock. The spectral and temporal properties of 
the collimated structures argue that they are focused jets of high speed 
material, as observed in the Crab. Such observations have already begun 
to inspire new axisymmetric MHD models that predict similar features
(e.g., Komissarov \& Lyubarsky 2004\nocite{kl04}), 
and ongoing observational studies 
promise to further constrain and refine such models. In particular, the 
confining mechanism for jets is not well-understood; many jets display some 
amount of curvature, with the Vela pulsar jet being an extreme example 
in which the morphology is observed to change on timescales of months
\citep{ptk+03}. 
This may be the result of pinch instabilities disrupting the toroidal
structure of the confining magnetic field (Begelman 1998), or could be 
indicative of an interaction of the jet material with the ambient medium.
There also appears to be a wide variation in the 
fraction of spin-down energy channeled into the jets, ranging from roughly 
$2.5\times 10^{-5}$ for
PSR~J0205+6449 in 3C~58 to nearly $10^{-3}$ for PSR~B1509$-$58 based
upon their synchrotron spectra (Slane et al. 2004, Gaensler et al. 2002). 
And, while Doppler 
beaming is invoked to explain the large brightness variations in jets and 
the associated counterjets, as well as around the observed toroidal 
structures, it is not clear that this alone is sufficient to explain the 
observed jet/counterjet brightness ratios. The nature of the curved and 
time-variable wisps outside 
the termination shock remains in question as well (see Section 3).

\chandra\ observations of G54.1+0.3 \citep{lwa+02} reveal
a central 136~ms pulsar \citep{clb+02} embedded 
in a diffuse
$1\farcm5 \times 1\farcm2$ ($2.2 {\rm\ pc} \times 1.8 {\rm\ pc}$)
nebula, assuming a distance of 5~kpc (Figure 2, center). 
The radio luminosity is among the lowest of the known young pulsars,
although pulsations at the radio period are clearly seen in X-rays as well.
The pulsar is surrounded by a $5\farcs7 \times
3\farcs7$ ($0.14 {\rm\ pc} \times 0.09 {\rm\ pc}$)
X-ray ring, with the long axis oriented roughly
north-south, suggesting an inclination angle of about $40^\circ$. The
X-ray emission is brightest along the eastern limb, and when interpreted
as the result of Doppler boosting, implies a post-shock velocity
of $\sim 0.6c$ (Lu et al. 2002; Romani \& Ng 2003\nocite{rn03}). 
The spectrum of the ring is
well-described by a power law and is harder than that for the surrounding 
diffuse emission. It is thus similar in many respects to the X-ray torus 
in the Crab Nebula. However, it comprises only $\sim 10\%$ of the extended 
emission in G54.1+0.3, while the torus in the Crab dominates the X-ray flux.
The pulsar in G54.1+0.3 is also much less luminous than the Crab Pulsar.

The \chandra\ data also reveal faint bipolar elongations running roughly
E-W, perpendicular to the long axis of the ring. The spectra from
these regions are also harder than that of the diffuse nebula, suggesting
lower synchrotron losses or more recent particle injection. These apparent
outflows, which presumably lie along the pulsar rotation axis, are more
diffuse than the jets in the Crab Nebula, yet appear to carry away a
considerably larger fraction of the energy; 
they comprise roughly the same luminosity as the central ring,
which is in stark contrast to the Crab where the torus outshines the jets
by a large factor. The structure of the elongation in the east appears
dominated by a clump of emission well-removed from the pulsar. 

\chandra\ observations of PSR~B1509$-58$ \citep{gak+02}
demonstrate that this young and energetic 
pulsar associated with G320.4--1.2 powers
an extended and extremely complicated PWN, with
structures on scales from $\sim10'$ 
($\sim 15$~pc at the distance of 5.2~kpc)
down to the spatial resolution limit (Figure 2, right).
On the largest scales, the elongated PWN has a clear axis
of symmetry centered on the pulsar, presumably representing the projected
orientation of the pulsar spin axis.  To the southeast of the
pulsar, the nebula is dominated by a narrow jet-like feature
approximately 6~pc in length, lying along this axis and displaying
a distinctly harder spectrum than the average for the nebula.
If this spectral difference is interpreted as a lack of synchrotron
cooling in the former component, one infers a
minimum flow speed in the jet-like component of $v_f \ga 0.2c$;
the lack of a similar feature to the north can be explained by Doppler
boosting for this speed of outflow if the pulsar's spin axis is inclined
to the line-of-sight by $\la 30^\circ$ (Gaensler et al., 2002).

In the central region of the PWN, a pair of semi-circular arcs lie $\sim0.5$ 
and $\sim1$~pc to the north of the pulsar. Gaensler et al. (2002) note that
if the inner region of these arcs represents the position
of the pulsar wind termination shock, then the flow time to the arcs
is much shorter than the synchrotron lifetime of the emitting particles
based on equipartition estimates of the magnetic field. Thus, unlike for
the Crab torus, where these timescales are similar, the emission from
the arcs is not the result of large synchrotron cooling at this position.
Instead, the arcs appear to resemble the series of concentric wisps seen 
for the Crab;
if interpreted as sites of electron compression in an ion-dominated flow
(as modeled for the Crab by Gallant \& Arons, 1994\nocite{ga94}), 
one can calculate an electron/positron
termination shock radius of $r_w \approx 0.5$~pc, and a nebular 
magnetic field of $B_n \sim8$~$\mu$G (Gaensler et al., 2002). 

The innermost region of 3C~58 (see Figure 3) consists of a 
bright, elongated 
compact structure centered on the pulsar J0205+6449. This inner nebulosity
is bounded along the western edge by a radio wisp 
\citep{fm93},
and is suggestive of a toroidal structure that is tilted about
a north-south axis, with the pulsar at its center, and with a jet-like
feature extending to the west. The eastern side of the toroid is slightly
brighter than the western side, suggesting that the eastern side is beamed 
toward us. If interpreted as a circular termination shock
zone, the inferred inclination angle in the plane of the sky is roughly
70 degrees \citep{shm02}.
The luminosity of the toroidal region is $L_x (0.5 - 10 {\rm\ keV})
= 5.3 \times 10^{33} {\rm\ erg\ s}^{-1}$.

\begin{figure}
\centerline{\epsfig{file=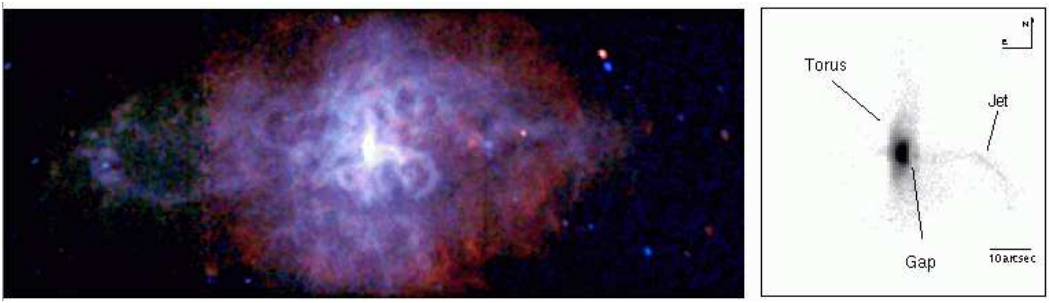,width=5.5in}}
\caption{
Left:
{\it Chandra} image of 3C 58. Colors correspond to the energy bands
0.5-1.0~keV (red), 1.0-1.5~keV (green), and 1.5-10 keV (blue).
Complex filamentary loops fill the interior region, and a softening of
the spectrum with radius is evidenced by the red outer regions -- an
effect resulting from both synchrotron aging of the electrons and the
presence of a soft thermal shell.
Right:
The innermost region of 3C~58 showing the neutron star embedded in an
elongated structure. A curved jet extends to the west, with a hint of a
counterjet component in the east.
}
\end{figure}

The elongated structure extending westward from the position of the pulsar has
the appearance of a jet (Figure 3, right). Its orientation is consistent, 
in projection, with the pulsar rotation axis inferred from the wind 
termination shock region discussed above, and also the east-west elongation 
of the entire PWN. The structure shows considerable curvature, similar to 
that seen in the Crab Pulsar jet. The power law index of the jet spectrum 
does not vary along the length of the feature, indicating that, as for 
PSR~B1509$-$58, the flow time 
across the jet is shorter than the synchrotron lifetime of the radiating 
particles. This sets a limit of $v > 0.01 c$ for the outflowing material 
assuming a minimum-energy magnetic field of $\sim 35 \mu{\rm G}$
\citep{shvm04}. A faint structure that may be a counterjet is observed to the
east of the pulsar. However, the jet/counterjet intensity ratio is $> 5$
which, from Eq. 2, requires $v > 0.8 c$, i.e., a supersonic flow.
The observed luminosity is
nearly a factor of 10 smaller than that for the torus.
For the Crab Nebula, the torus is nearly 20 times more luminous than the
jet in X-rays, while for PSR~B1509--58 the jet is brighter than the extended 
inner emission (Gaensler et al. 2002).

The jet/torus morphology observed in these PWNe provides the geometry of the
pulsar system, yielding both the projected direction of the spin axis and
the inclination angle. Modeling of such emission in other PWNe holds promise
for understanding the kicks that give pulsars their large space velocities
\citep{nr04}.
The jets observed in the Crab and Vela pulsars, for example, are 
aligned with their
proper motion vectors (Aschenbach \& Brinkman 1975\nocite{ab75}; 
Helfand et al. 2001\nocite{hgh01}).
If the kick that gave these pulsars their proper motion was
generated in the supernova explosion by some asymmetric mass ejection,
then this alignment requires an initial pulsar spin period that is
short relative to the kick timescale, so that the impulse of the kick
is averaged over many rotations of the star \citep{lcc01}.  
\cite{rn03} reach similar conclusions for PSR J0538+2817 in
the supernova remnant S147. By modeling the faint extended PWN emission
as a jet and torus, they derive a spin axis direction that is aligned with
the vector from the SNR center to the current pulsar position. 
For some pulsars [e.g. J0205+6449 in 3C~58 \citep{mss+02}
and J1811--1925 in G11.2--0.3 \citep{krv+01}],
we believe that the initial spin period was 
was relatively long $> 50$~ms.
If pulsar kicks are hydrodynamically-driven, where timescales of order
$100$~ms are expected \citep{lcc01}, this
would suggest that the proper motions of these pulsars should not necessarily 
be aligned with the jet direction.
Future radio timing observations of these pulsars
will ultimately lead to such proper motion measurements.  

\section{Time-Variable Structure in Pulsar Winds}

Early optical studies of the Crab revealed dynamic changes in the inner part 
of the nebula \citep{sca69}. 
Temporal variability in X-rays over long timescales was also revealed with
ROSAT observations \citep{ga99}.
Using a series of HST 
observations with short sampling intervals, \cite{hes98} showed 
that the Crab torus is comprised of
a series of wisps that move outward at speeds of $\sim 0.5$c, and concluded
that they result from unstable synchrotron cooling of magnetic flux tubes.
An alternative view is that the wisps are structures formed by compression
of the magnetic field and electron-positron pair plasma by magnetosonic waves 
produced by instabilities in ion gyration orbits, with the periodicity of
the waves being associated with the ion Larmor time 
(Gallant \& Arons 1994\nocite{ga94}; Spitkovsky \& Arons\nocite{sa04}), a 
picture that appears to be more consistent
with the structure of arcs observed around PSR~B1509$-$58 as well
(Gaensler et al., 2002). A series of joint HST and Chandra observations 
of the Crab shows
that these moving wisps are also observed in X-rays. The inner X-ray ring 
consists of a series of knots that routinely form, brighten, and then 
dissipate. This behavior is illustrated in Figure 4 which presents HST
and Chandra images of the inner Crab region at two different epochs; 
brightening of wisps to the northwest of the pulsar, as well as knots on
the southeastern portion of the inner ring, are particularly evident. In
addition, a feature at the base of the pulsar jet (indicated by arrows) is
seen to brighten in both the X-ray and optical bands between the two epochs.

\chandra\ observations of PSR B1509$-$58 also reveal compact knots just outside
the pulsar (Figure 2, left), and multi-epoch observations reveal variations
in both position and brightness of these knots \citep{gae04}. Similarly,
structures near PSR~J1811-1925 in G11.3$-$0.3 are observed to brighten and shift
in position between two widely-separated observations 
\citep{rtk+03}.
These features appear to correspond to unstable, quasi-stationary shocks in
the region just outside the wind termination shock, where the cold pulsar
wind is accelerated and joins the interior of the PWN.

Recent MHD models for PWNe created by anisotropic winds (e.g. Komissarov \&
Lyubarsky 2004\nocite{kl04})
have had success at reproducing the general jet/toroid
structure observed in the Crab and other systems. The termination shock
itself is much closer to the pulsar in the polar regions due to the toroidal
magnetic field geometry. Additional structure appears to result from a series 
of shocks, with the position of the observed radiation
depending on the geometry of the inner shock region as well as the viewing
angle of the observer. In particular, the model of Komissarov \&
Lyubarsky (2004) is able to reproduce the knot at the base of the Crab jet.
However, while this knot is located on the jet side of the pulsar axis,
knots observed in PSR B1509$-$58 are located on the opposite side. This
may imply a different viewing geometry, or may suggest that these variable
features correspond to a completely different mechanism. Additional study
of such dynamic knots are required to address this issue.

Additional complications that remain to be explained and modeled are the 
curved nature of many pulsar jets as well as the observed intensity variations
between jets and counterjets as well as around the toroidal structures. 
Additional high resolution observations are sure to provide new insights
and constraints into such modeling.

\begin{figure}
\centerline{\epsfig{file=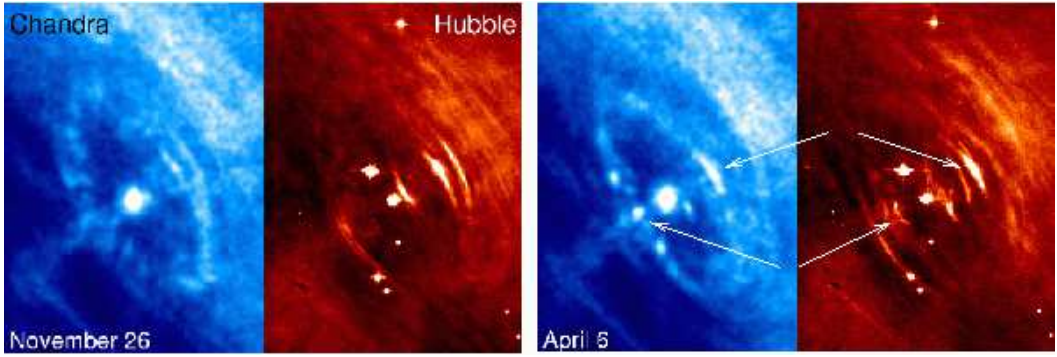,width=5.5in}}
\caption{
{\it Chandra} and HST images of the Crab Pulsar and its surroundings taken
at different epochs. The lower arrows in the image at the right indicate
compact structures at the base of the jet that have undergone dramatic 
changes in brightness between the two epochs. The upper arrows indicate 
regions just outside the wind termination shock that have undergone dramatic
changes in brightness.
}
\end{figure}

\section{Filamentary Structure in PWNe}

Extensive filamentary structure is observed in H$\alpha$, [OIII], and other
optical line images of the Crab Nebula. Based on their observed velocities, 
these filaments form an expanding shell of ejecta that surrounds the 
nonthermal optical emission from the nebula. High resolution
images with HST reveal detailed morphology and ionization structure 
suggesting that the filaments form from Rayleigh-Taylor instabilities
as the expanding relativistic bubble encounters slower moving ejecta
\citep{hss96b}, a picture supported by MHD 
simulations that show that 60-75\% of the swept-up mass ends up 
concentrated in such filaments (Jun 1998\nocite{jun98}, 
Bucciantini et al. 2004\nocite{bab+04}).
Radio observations reveal filaments that coincide with these optical 
filaments, presumably corresponding to synchrotron emission from regions 
of enhanced density and magnetic field in the form of magnetic sheaths that 
form as the pulsar-injected energy encounters the thermal filaments 
\citep{rey88b}.
Such filamentary structure is not observed in X-rays, however, 
suggesting that the electrons with sufficient energy to radiate X-rays do not 
reach the shell of filaments, or that in situ shock acceleration is not
producing them.
This is consistent with the observed smaller
extent of the X-ray emission in the Crab nebula relative to its radio size,
and indicates a larger magnetic field, and thus more rapid synchrotron
losses, than is observed in 3C~58 and
PSR~B1509$-$58.

Recent \chandra\ observations of 3C~58 reveal a complex of loop-like
filaments most prominent near the central regions of the PWN (Figure 3, left), 
but evident throughout the nebula \citep{shvm04}. 
These structures, 
whose X-ray spectra are nonthermal, are very well correlated with features 
observed in the radio band \citep{ra88}.
Optical observations reveal faint thermal filaments as well 
\citep{van78}, which presumably have an origin similar to that of the
Crab filaments. The velocities of these optical filaments in 3C~58
are $\sim \pm 900{\rm\ km\ s}^{-1}$ 
\citep{fes83}, sufficiently
high to indicate that the PWN is young, but too small to account for the
current size of \snr\ if the historical age is assumed -- one of
several standing problems with regard to its evolution. A detailed comparison
of the X-ray and optical images shows that most of the X-ray filaments do not 
have corresponding optical structures, however. While comparisons with 
deeper optical 
images are clearly needed, the fact that many of the X-ray features without 
optical counterparts are brighter than average in X-rays suggests that these 
may actually arise from a different mechanism. Slane et al. (2004) propose 
that the bulk of the discrete structures seen in the X-ray
and radio images of 3C~58 are magnetic loops torn from the toroidal field
by kink instabilities. In the inner nebula, the loop sizes are similar 
in diameter to
the size of the termination shock radius ($\sim 0.1$~pc), 
as suggested by Begelman (1998).
As the structures expand, they enlarge slightly as a consequence of the
decreasing pressure
in the nebula. Some of the observed X-ray structure in the outermost regions
may be the result of thermal filaments produced by Rayleigh-Taylor
instabilities, similar to the filaments in the Crab Nebula. An outer shell
of thermal X-ray emission (shown in red in Figure 5) demonstrates the presence 
of ejecta in these outer regions. While some of the optical
filaments do appear to be located in the central regions, these may lie
primarily along a shell seen in projection.

We note that considerable loop-like filamentary structure is evident in
\chandra\ observations of the Crab Nebula \citep{wht+00}. These
features are primarily observed encircling the bright Crab torus, perpendicular
to the toroidal plane, and may result from currents within the torus itself.
It is at least conceivable that such currents are signatures of the kink
instabilities suggested above.

\section{Conclusions}

High resolution X-ray studies of young pulsars and their associated wind
nebulae reveal a broad range of structures that are changing our view
on how pulsars interact with their surroundings. The clear evidence of
jets and toroidal structures gives a determination of the system geometry
as well as physical scales and luminosities for sites of particle acceleration.
This, in turn, constrains models for the overall morphology of PWNe as well
as for jet confinement, pulsar kicks, and the conversion of the cold pulsar
wind into the observed synchrotron nebulae. Filamentary structure provides
evidence of the ejecta into which the nebulae are expanding, and also
indications of magnetic structures whose nature is currently not well 
understood. While the Crab pulsar and its nebula have dominated theoretical
considerations on PWNe for the past thirty years, current observations
are providing a vast amount of information on similarities as well as gross
differences between the Crab and other PWNe, offering great promise of
dramatically expanding our understanding of these systems.

\section*{Acknowledgments}

I would like to thank the many colleagues with whom I have worked on studies
of young neutron stars and their wind nebulae for their contributions to my
understanding of numerous topics in this field. Special thanks to Okkie de
Jager and Rob Petre for their helpful comments in refereeing this paper.
This work was supported in
part by the National Aeronautics and Space Administration through contract 
NAS8-39073 and grants GO0-1117A and NAG5-9281.





\end{document}